\documentstyle[prb,preprint,aps]{revtex}    
\begin{document}
\draft
\newcommand{\beq}{\begin{equation}}
\newcommand{\eeq}{\end{equation}}
\newcommand{\beqa}{\begin{eqnarray}}
\newcommand{\eeqa}{\end{eqnarray}}

\title{Self-consistent current-voltage characteristics\\
of normal-superconductor interfaces}

\author{J. S\'anchez Ca\~nizares and F. Sols}
\bigskip
\address{
Departamento de F\'{\i}sica de la Materia Condensada, C-XII\\
Universidad Aut\'onoma de Madrid, E-28049 Madrid, Spain}

\maketitle
\begin{abstract}
We study
the nonlinear transport
properties of NS (normal-superconductor) and NSN structures
by means of a self-consistent microscopic description.
A nonzero superfluid velocity causes the
various quasiparticle
channels within S to open at different
voltages. The gap reduction
is very sensitive to the details of quasiparticle scattering.
At low temperatures, superconductivity,
some times in a peculiar gapless form,
may survive up to voltages much higher that $k_B T_c/e$.
The minimum voltage for quasiparticle transmission
is shown to decrease strongly with temperature
and with the transmittivity of the barrier.
\end{abstract}
\vspace{.5cm}

\pacs{PACS numbers: 74.40.+k, 74.50.+r, 74.90.+n}

\vspace{.5cm}
\narrowtext

During the last few years, there has been a renewed interest
in the transport properties of structures that involve both normal
and superconducting elements. This research has been largely
motivated by the study of mesoscopic
superconductivity, whose characteristic feature is
the phase-coherent propagation
of quasiparticles.
Due to its fundamental character, the
normal-superconductor (NS) interface has been the object of
especial attention, and a number of transport anomalies
have been associated to the existence of phase-coherent Andreev
reflection.\cite{ns}
These investigations
have been accompanied
by a revived interest in related questions
such as
the role of
self-consistency in microscopic descriptions based on the
resolution of the
Bogoliubov -- de Gennes equations,\cite{dege66}
\beqa
\left[ \begin{array}{cc}
H_0 & \Delta \\
\Delta^* & -H_0^* \end{array} \right]
\left[ \begin{array}{c} u_n \\ v_n \end{array} \right]
= \epsilon_n
\left[ \begin{array}{c} u_n \\ v_n \end{array} \right].
\eeqa
Here,
$H_0$ is the one-electron Hamiltonian, $\Delta$
is the gap function, and $(u_n,v_n)$ and
$\epsilon_n$ are, respectively,
the normalized wave function components and the energy of the
quasiparticle $n$. Self-consistency requires
\beq
\Delta = g \sum_{n} u_n v_n^* (1 - 2 f_n),
\eeq
where $g$ is the coupling constant and
$\{f_n\}$ are the occupation probabilities.
The implementation of
self-consistency is especially important in transport studies, since
it guarantees the conservation of electric
current.\cite{furu91,bagw94,sols94}
In a perfect one-dimensional superconductor, uniform
current-carrying solutions are of the form,\cite{dege66}
\beq
\Delta(x)=|\Delta|e^{i2qx}.
\eeq
In equilibrium,
$|\Delta|$ is a decreasing function of
$q$, with a critical current above which
$|\Delta|\!=\!0$.\cite{dege66,bagw94,bard72}
Self-consistency is  also
essential to understand the crossover
from the Josephson effect to bulk superconducting
flow.\cite{sols94}
A smooth crossover is possible thanks to the existence
of uniform
and solitonic (with a local gap depression)
configurations.\cite{sols94,mart94}
For scattering structures,
the use of scattering quasiparticle
channels compatible with (3) is sufficient to
guarantee asymptotic self-consistency.
\cite{sols95,sanc95}

The studies mentioned above either neglect self-consistency
or assume equilibrium of quasiparticles. These may be
reasonable assumptions in specific contexts,
namely, at low voltages (assumption of equilibrium)
or for currents much smaller than the critical current
of the perfect superconductor (neglect of self-consistency).
However,
a theoretical description that includes both self-consistency and
nonequilibrium effects is mandatory in transport problems involving
high applied voltages and large superconducting current densities.
In this Letter, we make
a first step in this direction.
First, we generalize
the classic scattering study by Blonder, Tinkham, and Klapwijk
\cite{blon82} on the NS interface
by including the effect of self-consistency.
We will see that this inclusion introduces major
changes in the I-V characteristics, but we will also
argue that
the most spectacular of them
are not observable in practice
because of the ultimate presence of a second
normal lead in any real experiment.\cite{lamb91}
For this reason, we
extend our analysis of the NS interface to the study of a NSN
structure. We compound the two interfaces by assuming
incoherent multiple scattering by the two barriers. This
simplification has the advantage that it permits us to focus
on the properties arising specifically
from the combination of self-consistency
and the superposition of two interfaces. In this
sense, we expect our predictions to be even more robust than
those we would have inferred from coherent scattering descriptions,
since these might be dependent on fine, not easily reproducible
details of the scattering
process. Therefore, the
realization of the physics discussed in this paper
does not require extremely small structures nor very low temperatures.
The only essential requirement is that, due to the
application of high voltages,
and in the absence of dilution effects,
the superconductor is forced to carry large currents.
The possible role of impurities
will be analyzed in a future paper.\cite{sanc95}

Instead of using a locally self-consistent description,
we have assumed that
the gap is of the form (3) everywhere
in the superconductor and zero in the normal lead.
This assumption
simplifies the numerics considerably and underlines
{\it the main effect of self-consistency}, which {\it is the
introduction of a nonzero superfluid velocity
$v_s\! \equiv \!\hbar q/m$ in the current-carrying configurations}.
In our scattering approach, we assume that
quasiparticles
are sent through the incoming channels of the semi-infinite
leads with chemical potentials that differ by
$eV$. Then,
we look for self-consistent
values of $|\Delta|$ and $q$ satisfying
Eqs. (1-3) and asymptotic current conservation.
We wish to emphasize that
{\it the occupations} $\{f_n\}$ {\it are those of a nonequilibrium
population}.
The existence of a nonzero $v_s$
has the major implication that
the energy thresholds for propagation across the
superconductor are modified, as shown in the insets of Fig. 1b.
The minimum energy to excite
a quasiparticle in the positive ($+k_F$) and negative
($-k_F$) branches becomes $\Delta_+$ and $\Delta_-$,
respectively, with
$\Delta_{\pm} \!\equiv\! |\Delta| \! \pm \! \hbar v_F q$.
$|\Delta|$ can in turn be supressed by current,
in a manner that is very sensitive to the
details of the scattering process.

At zero temperature,
electrons with energy $0\!<\!E\!<\!eV$ are injected
from the normal side onto the NS interface.
At low voltages
($eV \! < \! \Delta_-$), normal and Andreev reflection (AR) are the
only scattering mechanisms. In a AR process,
a hole is reflected on the N side and two electrons are transmitted to
the superconductor forming a Cooper pair.
Thus, one AR event
contributes a charge $2e$ that is transported by
the condensate.
As the voltage is raised, $q$ increases and,
when $eV \!>\! \Delta_-$,
Andreev transmission (AT) becomes also possible: one
electron is transmitted to the superconductor,
forming a Cooper pair and
creating a quasi-hole.
The amplitude for AT is proportional to $r^*t$,
where $r$ ($t$) is the
one-electron reflection (transmission)
coefficient.\cite{blon82}
One AT event
contributes a net charge of $e$ to the
current, $2e$ going to the condensate
and $-e$ to the quasiparticle part.
Thus, AT is less efficient than AR in that,
for the same contribution to the total current, it displaces
the condensate twice as much and thus costs more free energy.
We remark that there is
a range of voltages
in which AT is the {\it only} quasiparticle
transmission
mechanism. By contrast, in
a non self-consistent description, the AT and
normal transmission (NT) channels have the
same voltage threshold.\cite{blon82}

As $V$ and $q$ continue to increase,
the threshold $\Delta_-$ becomes negative.
In some cases, and especially at low temperatures,
a type of {\it gapless superconductivity} (GS)
with peculiar properties may survive.
As shown in the right inset of Fig. 1b,
the quasiparticles that are hidden below the
$\epsilon\!=\!0$ line in the
negative branch reemerge with positive energy in the opposite
branch.
For each quasiparticle $n$ that satisfies Eq. (1), there
is another solution $n'$ with energy
$\epsilon_{n'}\!=\!-\epsilon_n$
and wave components $u_{n'}\!=\!-v_n^*$ and
$v_{n'}\!=\!u_n^*$.\cite{dege66}
Creating $n',\sigma$ is equivalent to
destroying $n,-\sigma$ ($\sigma$ is the spin),
and
one is free to employ the $n$ or $n'$
description.\cite{sanc95}
The conventional choice is that which makes the energy positive, so
that quasiparticles are rare
at low temperatures. Following this convention,
a quasiparticle $n$ that acquires negative energy when
GS is reached must pass to be
described by its $n'$ counterpart. In the GS regime, low
energy electrons
coming from the N side can be normally
transmitted as a
quasi-electron in the lower positive branch.
Being of the $n'$-type,
this quasiparticle state has some unusual properties. Since
$u_nv_n^*\!=\!-u_{n'}v_{n'}^*$, it is clear from Eq. (2) that
the presence of $n'$-type
quasiparticles tends to {\it cancel} the effect
of conventional ($n$-type) quasiparticles.
In a standard scenario without
GS, the occupation
of a quasiparticle state tends to decrease the value of the gap
because the factor $(1-2f_n)$ changes sign and tends to
cancel the contribution from unoccupied states. In the presence
of GS,
the group of $n'$-type quasiparticles acts in the opposite
way. In front of already existing $n$-type quasiparticles,
these unconventional quasiparticle states
{\it tend to depress
the gap if they are empty and contribute
to reinforce superconductivity if they are occupied}.

In the spirit of Ref.\cite{blon82},
we have modeled the NS interface with
a step function gap and a one-electron potential barrier
$V_0 \delta(x)$ of effective strength $Z\!\equiv\!
V_0/\hbar v_F$.
The self-consistent current-voltage characteristics is
computed for different values of $Z$.
Except for a final prediction, we
focus on the
zero temperature case.
We remark that
the self-consistent solutions with
$|\Delta| \! \neq\! 0$
have been checked to
have {\it lower free energy} than the
trivial solutions with zero $q$ and $|\Delta|$.
We have taken a bandwidth
of $E_F\!=\!5$ eV, a cutoff energy of
$\hbar \omega_D \!=\! 0.1$ eV for Eq. (2),
and a zero temperature, zero current gap of
$\Delta_0\!=\!1$ meV.
This yields a critical temperature of $T_c\!=\!6.6$ K.
In Fig. 1a, we show the self-consistent I--V
characteristics
for an NS interface ($e\!>\!0$).
Non self-consistent results \cite{blon82}
are also shown for
comparison.

A detailed understanding of the physics behind the
I--V curves shown in Fig. 1a would require information
on the dependence of $|\Delta|$, $v_s$, and the
quasiparticle current on the applied voltage, which
will be presented elsewhere.\cite{sanc95}
Here, we only wish to note that, as the voltage
increases, the first rise in the current is due to
the opening of the AT channel (except in the $Z\!=\!0$
case, where it is forbidden because $r\!=\!0$).
The sharp decrease in $I$ for $Z\!=\!0$ and 0.5 is
due to the onset of the GS regime, where normal
quasiparticle transmission into the unconventional
branch begins to dominate, bringing down the value
of the conductance from close to 2 (in units of
$\!2 e^2/h$), typical of a transmissive NS interface,
to near unity, as corresponds to reflectionless normal
transport. This rather spectacular feature is however
not likely to be
measurable in practice,
since it would be destroyed by the presence
of a second normal lead.
For example, in the transparent case, this second
interface must send the value of the low $V$ differential
conductance back to its natural limit of 1.
On the other hand, nonequilibrium
effects that depend essentially on the total depopulation
of incoming channels from the S side can hardly be manifest
when a second normal lead sends quasiparticles
from the opposite side. These considerations motivate our
study of the NSN structure, for which
we assume identical barriers
and incoherent multiple scattering.
By symmetry, the potential
difference will be $V/2$ at each interface,
and the
flux of incoming electrons from the left N lead will be
mirrored by the flux of incoming holes from the right N lead.

In Fig. 1b, we show the resulting $I(V)$ curve for a NSN
structure. Non self-consistent results obtained with analogous
scattering assumptions are also shown.
Comparison with Fig. 1a reveals indeed major changes.
If $Z\!=\!0$, the presence of a superconducting segment
does not affect the conductance,
since it adds
nothing to perfect transmission.
Being $r\!=\!0$, AT is inhibited.
When $v\!\equiv eV/\Delta_0 \!=\!2$, the system
reaches GS, but,
remarkably,
$|\Delta|$ can be shown to remain unaffected.\cite{sanc95}
This happens because, when the unconventional branch
emerges, {\it its
two states are filled with very high probability and thus do not
contribute to depress the gap}.
This surprising behavior contrasts markedly
with that expected for situations in which quasiparticles
are in equilibrium. In such cases, $|\Delta|$ is
predicted to decrease very quickly as $\Delta_-$
becomes negative \cite{bagw94,bard72}. Here,
a strong departure from equilibrium gives rise to
an effective translational invariance
that explains the small sensitivity of $|\Delta|$
to high voltages.

For thicker barriers, we see that, in a non
self-consistent calculation, the jumps in the
current are much smoother and occur at higher $V$.
For $Z\!=\!0.5$,
the AT channel opens for $v\!\simeq\!
1.3$, causing a small decrease in $|\Delta|$\cite{sanc95},
and GS sets in for $v\!\simeq\!
1.8$.
The onset of GS
is almost inconsequential for the gap
and total current behavior,
for reasons similar to those of
the transparent case.
We conclude that, for NSN structures with low $Z$,
the voltage is inefficient in
destroying superconductivity, although this may
remain quite marginal (in the GS regime, the actual
current carried by the condensate is a small
fraction of the total current).

For $Z\!=\!1$ and 2, the system
bypasses the only-AT regime and
jumps directly to GS
for $v\!\simeq\!1.7$ and $v\!\simeq\!1.9$,
respectively.
We attribute this behavior to the
energetic cost of opening two
AT channels.
The superconductor finds more favourable to carry charge through
the NT channel by going gapless.
An important effect is that multiple
normal reflection increases the residence time of the injected
quasiparticles. At high $Z$, one can prove by
invoking unitarity \cite{sanc95} that all the effective occupations
within S tend to 1/2 (two of the four incident channels are
empty), as opposed to the low $Z$ case (see above)
in which conventional
(unconventional) channels are occupied with
low (high) probability.
This effect causes
$|\Delta|$ to decrease with increasing $Z$, in marked
contrast with the behavior
found for the NS case.\cite{sanc95}

The above discussion shows that, for the values of $Z$ considered,
superconductivity can exist
at least up to voltages of order $5k_BT_c/e$.
Internal thermalization of
quasiparticles would tend
to destroy this effect. In this sense, we can state
that {\it the survival of superconductivity
at high voltages is a
nonequilibrium effect.}

We would like to end this discussion with a few comments about the
effect of temperature and dimensionality. Provided
the superconductor can still be treated as
quasi--one-dimensional (width much smaller
than coherence and penetration lengths),
the effect of many channels is that of
requiring higher $q$'s to observe similar effects,
since high-lying channels have a smaller effective
longitudinal Fermi velocity. Also, the presence
of many channels effectively breaks the translational
invariance we found for very transmissive barriers.
\cite{sanc95}
As a consequence, the survival of superconductivity
at high voltages is considerably weakened, in
accordance with the behavior found in semiclassical
nonequilibrium superconductivity.\cite{gray}
Apart from smoothening the structure of the $I(V)$ curves,
the main effect of temperature
is that of bringing
down the scale of gap energies. With smaller gaps, the condensate
needs higher values of $q$ and the result is that self-consistency
effects are even more important. A major consequence is that,
as the voltage increases, the thresholds for the novel
regimes discussed above are anticipated.
Fig. 2 shows that the voltage threshold
for the onset of AT, $V_{AT}$,
which determines the first peak in the differential
conductance, is a decreasing function of temperature,
even when measured in units of the zero current gap $\Delta_0(T)$.
One may also note that $V_{AT}$ increases with $Z$,
due to a global
reduction of $v_s$.

In conclusion, we have performed a self-consistent calculation
of the nonlinear transport properties of NS and NSN structures.
A major effect of self-consistency
is that of forcing the current-carrying superconductor to
have a nonzero superfluid velocity.
This causes the
various quasiparticle channels to enter into action at different
voltages. The study of the NS interface has served to identify
the scattering processes separately and has allowed us to
understand in detail the properties of the more realistic
NSN system.
The gap amplitude is depressed
by the flow of current in a way that
is very sensitive to the details
of the scattering problem.
An important nonequilibrium effect is that,
at low temperatures, a type of gapless
superconductivity may
survive up to voltages considerably greater than $k_BT_c/e$,
although this effect may be weakened in the presence of
many channels.
The onset of Andreev trasmission
is signaled by a peak in the differential
conductance whose position relative to the zero current gap
has been predicted to decrease with temperature and with the
transmittivity of the barrier.

{\it Note added.} After this work had been completed, we
learned about the content of Ref.\cite{lamb95}, where a
related NSN model has been studied.
In Ref. \cite{lamb95}, one interface is always transmissive
and multiple phase-coherent
scattering is  assumed. In spite of these differences,
the model studied by Martin and Lambert \cite{lamb95}
displays some common robust features like
the splitting of voltage thresholds and the depression
of the gap.

\acknowledgments

We wish to thank Carlo Beenakker, Jaime Ferrer, Colin Lambert,
and Gerd Sch\"on
for valuable discussions.
This project
has been supported by Direcci\'on General
de Investigaci\'on Cient\'{\i}fica
y T\'ecnica, Project no. PB93-1248, and by
the Human and Capital Mobility Programme of the EEC.
J.S.C. acknowledges support from Ministerio de
Educaci\'on y Ciencia through a FPI fellowship.

\begin{figure}
\caption{
Solid lines show the self-consistent I-V characteristics
of a NS (a) and a NSN (b) structure,
for four values of $Z$. Dotted lines show the results
from a non self-consistent calculation (see Ref. 9).
The dashed line in (a) gives the NN result.
Insets: schematic
quasiparticle dispersion relation, $\epsilon(k)$, in the (a) normal
and (b) superconducting leads. Right inset in (b)
shows gapless superconductivity. Filled
(empty) circles indicate electron- (hole-) like propagation.
}
\end{figure}

\begin{figure}
\caption{
Position of the first peak in the differential conductance
(caused by the onset of Andreev transmission),
in units of the zero current gap, plotted as function of temperature
for $Z\!=\!0.5$, 1, and 2.
}
\end{figure}

\end{document}